\documentclass[letterpaper]{JHEP3}

\def\sech{\mathop{\mathrm{sech}}\nolimits}

\title{Absorption cross section in warped AdS$_3$ black hole revisited}

\author{Hsien-Chung Kao\\
Department of Physics, National Taiwan Normal University,
Taipei 116, Taiwan\\
\email{hckao@ntnu.edu.tw}
}

\author{Wen-Yu Wen\\
Department of Physics and Center for Theoretical Sciences \& \\ 
Leung Center for Cosmology and Particle Astrophysics,\\
National Taiwan University, Taipei 106, Taiwan\\
\email{steve.wen@gmail.com}}

\abstract{We investigate the absorption cross section for minimal-coupled scalars in the warped AdS$_3$ black hole.  According to our calculation, the cross section reduces to the horizon area in the low energy limit as usually expected in contrast to what was previously found.  We also calculate the greybody factor and find that the effective temperatures for the two chiral CFT's are consistent with that derived from the quasinormal modes.  Observing the conjectured warped AdS/CFT correspondence, we suspect that a specific sector of the CFT operators with the desired conformal dimension could be responsible for the peculiar thermal behaviour of the warped AdS$_3$ black hole.} 


\begin{document}

\section{Introduction and Summary}
It has been proved to be universal that for all spherically symmetric black holes the low energy cross section for massless minimally coupled scalars is always equal to the area of the horizon\cite{Das:1996we}.  Furthermore, if the energy dependence is retained, the greybody factor or decay rate can also be obtained from the corresponding conformal field theory of the black hole\cite{Das:1996wn,Maldacena:1996ix,Maldacena:1997ih,Cvetic:1997uw,Cvetic:1997xv,Gubser:1996zp,Birmingham:1997rj}. Making of the greybody factor, we can correctly reproduced the Bekenstein-Hawking entropy of the black hole. Recently, a black hole solution in the topological massive gravity (TMG) with a negative cosmological constant was constructed\cite{Bouchareb:2007yx}. It was soon realized that the black hole can be viewed as discrete quotients of the warped  AdS$_3$ spacetime \cite{Anninos:2008fx} just like the BTZ black hole as discrete quotients of the AdS$_3$.  Following the conventions in \cite{Anninos:2008fx}, the metric of the warped AdS$_3$ black hole is given by 
\begin{eqnarray}\label{warped_AdSBH}
&&ds^2=-N^2(r)dt^2+\ell^2R^2(r)[d\phi+N_\phi(r)dt]^2+\frac{\ell^4 dr^2}{4R^2(r)N^2(r)},\\
{\rm with}&&R^2(r)\equiv\frac{r}{4}\big( 3(\nu^2-1)r+(\nu^2+3)(r_++r_-)-4\nu \sqrt{r_+r_-(\nu^2+3)} \big), \nonumber\\
&&N^2(r)\equiv\frac{\ell^2(\nu^2+3)(r-r_+)(r-r_-)}{4R^2}, \nonumber\\
&&N_\phi(r)\equiv\frac{2\nu r-\sqrt{r_+r_-(\nu^2+3)}}{2R^2}, \nonumber
\end{eqnarray}
and the horizon area 
\begin{equation}
{\cal A}_H = \pi\left\{2\nu r_+ -\sqrt{r_+r_-(\nu^2+3)}\right\}.
\end{equation}
It is known that when $\nu =1$, it reduces to the BTZ black hole case.
It was conjectured in \cite{Anninos:2008fx} that this background has a holographic description in terms of two-dimensional conformal field theory with uneven 
temperatures in the left and right sectors, given by
\begin{eqnarray}\label{temperature}
&&T_H^{-1} = \frac{4\pi\nu}{\nu^2+3}\frac{T_L+T_R}{T_R},\nonumber\\
&&T_L = \frac{\nu^2+3}{8\pi}(r_++r_--\frac{\sqrt{r_+r_-(\nu^2+3)}}{\nu}),\nonumber\\
&&T_R = \frac{\nu^2+3}{8\pi}(r_+-r_-).
\end{eqnarray}
The corresponding central charges are
\begin{equation}
c_L = \frac{\ell}{G}\frac{4\nu}{\nu^2+3},\qquad c_R = \frac{\ell}{G}\frac{5\nu^2+3}{\nu(\nu^2+3)}.
\end{equation}
There have been several discussions on the thermodynamical properties of this black hole.  Among them, we remark some unusual features: Firstly, a computation of the low energy absorption cross section was carried out in \cite{Oh:2008tc} and to our surprise it did not reduce to the horizon area even in the BTZ limit\footnote{Although it was later shown in a separate paper \cite{Oh:2009if} that using a different coordinate system, absorption cross section did reduce to horizon area at this critical point.}.  This result seems to be in contradiction with an earlier statement in \cite{Das:1996we} since a $U(1)$ isometry is still preserved in the warped geometry, though the area has been rescaled by deformation.  Secondly, the quasinormal modes of this background were discussed in \cite{Chen:2009rf} and their dispersion relation was shown no longer linear in contrast to what is expected from the usual AdS/CFT correspondence.  When $\nu>1$, a $\omega^2$ term is included in the definition of the conformal dimension of the operator which is coupled to the bulk scalar field thanks to the deformation.  It was argued in the conclusion that the conventional AdS/CFT correspondence only made sense for very small deformation, in other words, $3(\nu^2-1)\omega^2 \ll m^2(\nu^2+3)$ for scalar mass $m$.  In this paper, we would like to investigate again these peculiar features more carefully.  We find a disagreement with Ref. \cite{Oh:2008tc}.  More specifically, our computation shows that the low energy absorption cross section does reduce to the horizon area, and therefore the statement in Ref. \cite{Das:1996we} still holds.  In addition, we obtain the desired greybody factor,
\begin{equation} 
\sigma_{abs} \propto \frac{(2h^*_+-1)\sinh\left(\frac{\omega}{2T_H}\right)}{\omega \Gamma^2(2h^*_+)}\bigg| \Gamma\left(h^*_++i\frac{\omega}{4\pi \tilde{T}_R}\right) \Gamma\left(h^*_++i\frac{\omega}{4\pi \tilde{T}_L}\right)\bigg|^2.
\end{equation}
Here, the effective temperatures are given by \cite{Chen:2009rf}  
\begin{eqnarray}
&&\hskip -1cm \tilde{T}_L = \frac{T_L}{\delta} = \frac{\nu^2+3}{8\pi\nu};\\
&&\hskip -1cm \tilde{T}_R = \frac{T_R}{\delta} = \frac{(\nu^2+3)(r_+-r_-)}{8\pi(\nu(r_++r_-)-\sqrt{r_+r_-(\nu^2+3)})}\\
{\rm with}&&\delta \equiv \nu(r_++r_-)-\sqrt{r_+r_-(\nu^2+3)}, \nonumber
\end{eqnarray}
and the conformal dimension
\begin{equation}
h^*_+ = \frac{1}{2}\bigg(1+ \sqrt{1-\frac{12(\nu^2-1)}{(\nu^2+3)^2}\omega^2+\frac{4m^2\ell^2}{\nu^2+3}} \bigg).
\end{equation}
We also carry out the computation of massless scalar field in the (warped) AdS/CFT correspondence and conjecture that in the dual conformal theory living on the $(t,u)$-plane there exists a specific operator ${\cal O}_*$ with the desired dimension which is responsible for the peculiar thermal behaviour found from the quasinormal modes.  

One evidence supporting this statement is that the greybody factor can be correctly reproduced by using correlator built from ${\cal O}_*$, that is
\begin{equation}\label{deformed_correlator}
\bigg< {\cal O}_*(t,u){\cal O}_*(t',u') \bigg> \sim \frac{e^{\pm i\frac{2\nu\omega}{\nu^2+3}(u-u')}}{|t-t'|^{2h^*_+}}.
\end{equation}
The structure of this paper is outlined as follows:  in section 2 we review the derivation of the differential equation for a massive scalar field probing the AdS$_3$ black hole background.  In section 3, we discuss the subtlety arising from which one of the two limits is taken first, the massless or the low energy limit.  In section 4, we revisit the computation of the absorption cross section and then obtain the greybody factor in section 5.  In section 6, we attempt to derive the same greybody factor from the perspective of the effective conformal field theory.  In section 7, we have discussion on the superradiant modes.  For completeness, we also include some discussion on the warped AdS$_3$/CFT$_2$ correspondence in the Appendix.

\section{Scalar field in warped AdS$_3$ black hole}
To set up the notation for later computation, we will first repeat the derivation of the governing differential equation describing a massive scalar field probing the background described in eq. (\ref{warped_AdSBH}) \cite{Oh:2008tc,Chen:2009rf,Chakrabarti:2009ww}.  The Klein-Gordon equation in a curved background is given by
\begin{equation}
\left(\frac{1}{\sqrt{-g}}\partial_\mu \sqrt{-g}\partial^\mu-m^2 \right)\Phi=0.
\end{equation}
Using separation of variables
\begin{equation}
\Phi(t,r,\theta)=e^{-i\omega t +i\mu\theta}\phi(r),
\end{equation}
we obtain the radial equation
\begin{eqnarray}\label{ode}
&& \hskip-1cm \frac{d^2 \phi(r)}{dr^2}+\frac{2r-r_+-r_-}{(r-r_+)(r-r_-)}\frac{d\phi(r)}{dr}-\frac{(\alpha r^2+\beta r+\gamma)}{(r-r_+)^2(r-r_-)^2}\phi=0,\\
{\rm with} &&\alpha=-\frac{3\omega^2(\nu^2-1)}{(\nu^2+3)^2}+\frac{m^2\ell^2}{\nu^2+3},\nonumber\\
&&\beta=-\frac{\omega^2 (\nu^2+3)(r_++r_-)-4\nu\left[\omega^2\sqrt{r_+r_-(\nu^2+3)}-2\mu\omega\right]}{(\nu^2+3)^2}+\frac{m^2\ell^2(r_++r_-)}{\nu^2+3},\nonumber\\
&&\gamma=-\frac{4\mu\left[\mu-\omega\sqrt{r_+r_-(\nu^2+3)}\right]}{(\nu^2+3)^2}+\frac{m^2\ell^2r_+r_-}{\nu^2+3}.\nonumber
\end{eqnarray}
It is convenient to introduce the coordinate
\begin{equation}
z=\frac{r-r_+}{r-r_-}
\end{equation}
such that the outer horizon is situated at $z=0$ and spatial infinity at $z=1$.  The above equation then takes the form
\begin{eqnarray}
&& \hskip-1cm z(1-z)\phi''(z)+(1-z)\phi'(z)+\left[\frac{A}{z}+\frac{B}{1-z}+C\right]\phi(z)=0;\nonumber\\
{\rm with} &&A=\frac{4(\omega\Omega_+^{-1}+\mu)^2}{(r_+-r_-)^2(\nu^2+3)^2},\nonumber\\
&&B=-\alpha,\nonumber\\
&&C=-\frac{4(\omega{\Omega}_-^{-1}+\mu)^2}{(r_+-r_-)^2(\nu^2+3)^2},
\end{eqnarray}
with prime denoting differentiation with respect to $z$.
$\Omega_+$ and $\Omega_-$ are the angular velocity at the outer and inner horizon respectively, and
\begin{equation}
\Omega_+^{-1}=\nu r_+-\frac{\sqrt{r_+r_-(\nu^2+3)}}{2},\qquad \Omega_-^{-1}=\nu r_--\frac{\sqrt{r_+r_-(\nu^2+3)}}{2}.
\end{equation}
Upon removing the poles in the last term of the above equation through the following ansatz
\begin{eqnarray}\label{conformal_weight}
&&\hskip-1cm \phi(z)=z^p(1-z)^q u(z),\\
{\rm with}&&p= -i\sqrt{A}, \qquad q=\frac{1}{2}\left(1-\sqrt{1+4\alpha}\right),\nonumber
\end{eqnarray}
we reach a standard form of the hypergeometric differential equation
\begin{eqnarray}\label{hypergeometric}
&&\hskip-1cm z(1-z)u''(z)+\{c-(a+b+1)z\}u'(z)-ab u(z)=0,\\
{\rm with}&&a=p+q+\sqrt{C}, \qquad b=p+q-\sqrt{C},\qquad c=2p+1.\nonumber
\end{eqnarray}
It is well known that the two independent solutions are given by
\begin{eqnarray}
&&u_1(z)={}_2F_1(a,b;c;z),\\
&&u_2(z)=z^{1-c}{}_2F_1(a-c+1,b-c+1;2-c;z).
\end{eqnarray}
We comment that these solutions are standard in $z$ coordinate, which agrees with those in Refs. \cite{Chen:2009rf,Chakrabarti:2009ww}.  In Ref. \cite{Oh:2008tc}, the computation was carried out in the $r$ coordinate and a different set of independent solutions were obtained in equation (3.6) in their paper.  Although their solutions are equivalent to ours up to linear combination, we suspect that their choice could cause some complication when comparing solutions in the asymptotic region and therefore makes it difficult to see how the cross section reduces to the horizon area at low energy.

\section{Tortoise coordinate and effective potential}
It is instructive to see equation (\ref{ode}) in the tortoise coordinate $r^*$, such that
\begin{equation}
\phi^*(r^*) \equiv z(r) \phi(r), \qquad r^* = f(r).
\end{equation}
$f(r)$ and $z(r)$ are uniquely determined by requiring that eq. (\ref{ode}) can be brought into the following form:
\begin{eqnarray}
\left[-\frac{d^2}{dr^{*2}}-\omega^2+U^*(r^*)\right]\phi(r^*)=0.
\end{eqnarray}
We observe that in the spatial infinity, the effective potential 
\begin{eqnarray}
\lim_{r\to\infty}U^*(r)\to \frac{(\nu^2+3)(\nu^2+3+4m\ell^2)}{12(\nu^2-1)}.
\end{eqnarray}
In the case of the BTZ black hole where $\nu=1$, we always expect an infinite potential wall in the spatial infinity.  For a generic $\nu \neq 1$, however, it becomes a wall of finite height.  Therefore for $\omega^2<U^*(\infty)$ we may apply the Dirichlet boundary condition and the wavefunction vanishes in the spatial infinity. In contrast, for $\omega^2 \ge U^*(\infty)$ we expect the wavefunction to be nonvanishing there.  As we will see later, this new feature may bring subtlety when both the massless and low energy limits are taken.  

\section{Revisit absorption cross section}
To obtain the absorption cross section, we consider the scattering process that an in-going flux comes from the spatial infinity and interacts with the black hole. It is then partially reflected backwards as out-going flux to the spatial infinity and the rest absorbed into the black hole.  One way to achieve this goal is to consider a wavefunction with pure in-going boundary condition at the horizon and carefully decompose it into the in-going and out-going parts in the asymptotic region.  This determines the form of the wavefunction for the scalar field:
\begin{equation}\label{wave_function}
\phi = C_{in} z^{p}(1-z)^{(a+b-2p)/2}{}_2F_1(a,b;c;z), 
\end{equation}
where $p$ is given in eq. (\ref {conformal_weight}), $a,b,c$ in eq. (\ref{hypergeometric}) and $C_{in}$ an arbitrary coefficient.  Expanding this solution at small $z$, we obtain the asymptotic form near the outer horizon:
\begin{equation}
\phi_+ \simeq C_{in}\left(\frac{r-r_+}{r_+-r_-}\right)^{-i\frac{2\mu+\omega\left\{2r_+\nu-\sqrt{r_+r_-(\nu^2+3)}\right\}}{(r_+-r_-)(\nu^2+3)}} +\cdots
\end{equation}
Now let us consider the same wavefunction (\ref{wave_function}) near the spatial infinity:
\begin{eqnarray}\label{phi_infty}
\phi_\infty \simeq &&C_{in}\frac{\Gamma(c)\Gamma(c-a-b)}{\Gamma(c-a)\Gamma(c-b)}\left(\frac{r}{r_+-r_-}\right)^{-\frac{1}{2}+\frac{1}{2}\sqrt{1+4\alpha}}\left[1+\frac{ab}{1-c+a+b}\left(\frac{r}{r_+-r_-}\right)^{-1}\right] \\
&&+C_{in}\frac{\Gamma(c)\Gamma(a+b-c)}{\Gamma(a)\Gamma(b)}\left(\frac{r}{r_+-r_-}\right)^{-\frac{1}{2}-\frac{1}{2}\sqrt{1+4\alpha}}\left[1+\frac{(c-a)(c-b)}{1+c-a-b}\left(\frac{r}{r_+-r_-}\right)^{-1}\right]+\cdots \nonumber
\end{eqnarray}
On the other hand, we may also take the limit $r\to \infty$ in eq (\ref{ode}) first and it becomes
\begin{equation}
\phi''(r)+\frac{2}{r}\phi'(r)-\frac{\alpha r+\beta }{r^3}\phi(r)=0.
\end{equation}
It has two independent solutions
\begin{equation}
r^{\frac{-1}{2}}I_{-\sqrt{1+4\alpha}}\left(2\sqrt{\frac{\beta}{r}}\right), \qquad r^{\frac{-1}{2}}I_{\sqrt{1+4\alpha}}\left(2\sqrt{\frac{\beta}{r}}\right),
\end{equation}
where the modified Bessel function $I_n(x)$ has the following expansion at small $x$:
\begin{equation}
I_n(x) \simeq \frac{x^n}{\Gamma(1+n)}\left[ 1+\frac{\Gamma(1+n)}{2^2\Gamma(2+n)}x^2 + \cdots \right].
\end{equation}
We then define the in-going and out-going waves as a linear combination of the leading terms, i.e.
\begin{eqnarray}\label{phi_decomposition}
&&\phi_{\infty} \simeq \phi_{\infty}^{in}(r) + \phi_{\infty}^{out}(r),\nonumber\\
&&\phi_{\infty}^{in} = A_{in}\left(r^{-\frac{1}{2}+\frac{1}{2}\sqrt{1+4\alpha}}-\frac{i\eta}{\pi}r^{-\frac{1}{2}-\frac{1}{2}\sqrt{1+4\alpha}}\right),\nonumber\\
&&\phi_{\infty}^{out}= A_{out}\left(r^{-\frac{1}{2}+\frac{1}{2}\sqrt{1+4\alpha}}+\frac{i\eta}{\pi}r^{-\frac{1}{2}-\frac{1}{2}\sqrt{1+4\alpha}}\right).\nonumber\\
\end{eqnarray}
Here, $\eta$ is some undetermined positive coefficient, which we take to be independent of $\omega$.  In comparison with eq. (\ref{phi_infty}), we read
\begin{eqnarray}\label{A_coeff}
&&A_{in} = \frac{C_{in}}{2}\left\{\frac{\Gamma(c)\Gamma(c-a-b)}{\Gamma(c-a)\Gamma(c-b)}(r_+-r_-)^{\frac{1}{2}(1-\sqrt{1+4\alpha})}+i\frac{\pi}{\eta}\frac{\Gamma(c)\Gamma(a+b-c)}{\Gamma(a)\Gamma(b)}(r_+-r_-)^{\frac{1}{2}(1+\sqrt{1+4\alpha})}\right\},\nonumber\\
&&A_{out}=\frac{C_{in}}{2}\left\{\frac{\Gamma(c)\Gamma(c-a-b)}{\Gamma(c-a)\Gamma(c-b)}(r_+-r_-)^{\frac{1}{2}(1-\sqrt{1+4\alpha})}-i\frac{\pi}{\eta}\frac{\Gamma(c)\Gamma(a+b-c)}{\Gamma(a)\Gamma(b)}(r_+-r_-)^{\frac{1}{2}(1+\sqrt{1+4\alpha})}\right\}.\nonumber\\
\end{eqnarray}
We are now ready to compute the in-going flux via the following definition:
\begin{equation}\label{AinAout}
{\cal F}=\frac{2\pi}{i}(r-r_+)(r-r_-)\left[ \phi^*(r)\phi'(r)-\phi(r)\phi'^{*}(r)\right].
\end{equation}
The absorption coefficient for the s-wave ($\mu=0$) reads,
\begin{equation}\label{absorp}
{\cal T}\equiv \frac{{\cal F}(\phi_+)}{{\cal F}(\phi_{\infty}^{in})}=\frac{\omega {\cal A}_H}{4\eta\sqrt{1+4\alpha}}\left|\frac{C_{in}}{A_{in}}\right|^2.
\end{equation}
Note that $A_{in}$ is generally quite complicated for arbitrary mass and angular frequency.  Here, we would like to focus on the low energy limit for massless scalar. Under such conditions, we find that the coefficient of the first term in both curly brackets in eq. (\ref{AinAout}), $\frac{\Gamma(c)\Gamma(c-a-b)}{\Gamma(c-a)\Gamma(c-b)}$, always reduces to 1 regardless of whether the low energy or the massless limit is taken first.  On the other hand, the coefficient of the second term 
\begin{equation}
\frac{\Gamma(c)\Gamma(a+b-c)}{\Gamma(a)\Gamma(b)}\to \left\{ 
\begin{array}{l l}
  0, & \quad \mbox{ if $\omega\to 0$ is taken first;}\\
  \frac{2(\Omega_+^{-2}-\Omega_-^{-2})}{3(r_+-r_-)^2(\nu^2-1)},  & \quad \mbox{ if $m\to 0$ is taken first.}\\
\end{array} \right.
\end{equation}
From these results, we can then calculate the cross section for a massless scalar at low energy 
\begin{equation}\label{cross_section}
\sigma^0_{abs}\equiv\lim_{\omega\to 0}\frac{{\cal T}}{\omega} = {\cal A}_H,
\end{equation}
if $\eta$ is chosen properly
\begin{equation}
\eta = \left\{
\begin{array}{l l}
 1, & \quad \mbox{if $\omega\to 0$ is taken first;}\\
 \frac{1\pm \sqrt{1-\frac{8\pi^2(\Omega_+^{-2}-\Omega_-^{-2})}{3(\nu^2-1)}}}{2}, & \quad \mbox{if $m\to 0$ is taken first.}\\
 \end{array} \right.
\end{equation}
We comment that this subtle difference arises from the distinct behaviour of the wavefunction in the spatial infinity thanks to the competition between $\omega^2$ and $U^*(\infty)$.  If the low energy limit is taken first, the wavefunction is vanishing in the spatial infinity.  If it is the other way around then the wavefunction would be nonvanishing there.

\section{Greybody factor}
In the above computation of the absorption cross section, if one retains the $\omega$ dependence instead of setting it to be identically zero, then one can calculate the greybody factor. This has been done for the BTZ black hole \cite{Birmingham:1997rj}, but it remains a nontrivial check for the warped AdS$_3$ black hole.  Following procedures similar to those taken to obtain the result in eq. (\ref{cross_section}), we find the relevant part of the absorption cross section for s-wave\footnote{It is straightforward to generalized to the case $\mu \neq 0$ by shifting terms in eq. \ref{greybody} such that $\frac{\omega}{4\pi \tilde{T}_R} \to \frac{\omega + \tilde{\mu}}{4\pi \tilde{T}_R}$ and $\frac{\omega}{2\pi T_H} \to \frac{1}{4\pi}(\frac{\omega+\tilde{\mu}}{\tilde{T}_R}+\frac{\omega}{\tilde{T}_L})$, where $\tilde{\mu}\equiv \frac{\mu}{(2\nu(r_++r_-)-\sqrt{r_+r_-(\nu^2+3)})}$.}:
\begin{eqnarray}\label{greybody}
&&\sigma_{abs} \propto \frac{1}{\sqrt{1+4\alpha}}\left| \frac{\Gamma\left[\frac{1}{2}(1+\sqrt{1+4\alpha})-i\frac{\omega}{4\pi \tilde{T}_R}\right]\Gamma\left[\frac{1}{2}(1+\sqrt{1+4\alpha})-i\frac{\omega}{4\pi \tilde{T}_L}\right]}{\Gamma\left[1-i\frac{\omega}{2\pi T_H}\right]\Gamma\left[\sqrt{1+4\alpha}\right]} \right|^2,
\end{eqnarray}
where $\alpha$ is given in (\ref{ode}) and satisfies the reality condition $1+4\alpha \ge 0$.  The {\it effective} temperatures are known to be related to the ones in (\ref{temperature}) by a proper coordinate transformation\cite{Chen:2009rf}:
\begin{eqnarray}
&&\tilde{T}_R = \frac{T_R}{(\sqrt{r_+}-\sqrt{r_-})^2} \bigg|_{\nu=1} = \frac{1}{2\pi}\frac{\sqrt{r_+}+\sqrt{r_-}}{\sqrt{r_+}-\sqrt{r_-}},\nonumber\\
&&\tilde{T}_L = \frac{T_L}{(\sqrt{r_+}-\sqrt{r_-})^2} \bigg|_{\nu=1} = \frac{1}{2\pi}.
\end{eqnarray}
First we would like to check that one does recover the results in Ref. \cite{Birmingham:1997rj} in the BTZ limit, i.e. $\nu\to 1$.  For simplicity, we will restrict our discussion to the massless case.  Since $\alpha\to 0$ in this limit, we have
\begin{eqnarray}
&& a\to-i\frac{\omega}{4\pi \tilde{T}_R},\qquad b\to-i\frac{\omega}{4\pi \tilde{T}_L},\nonumber\\
&& \frac{\Gamma(c)\Gamma(c-a-b)}{\Gamma(c-a)\Gamma(c-b)}\to \frac{\Gamma(1+a+b)\Gamma(1)}{\Gamma(1+b)\Gamma(1+a)}.
\end{eqnarray} 
Using the identity
\begin{equation}\label{identity_1}
|\Gamma(1-ia)|^2 = \frac{\pi a}{\sinh{\pi a}},
\end{equation}
we then recover the greybody factor
\begin{equation}
\sigma_{abs}\propto \omega \frac{e^{\omega/T_H}-1}{\left(e^{\omega/2\tilde{T}_L}-1\right)\left(e^{\omega/2\tilde{T}_R}-1\right)}.
\end{equation}

It can be seen that the above derivation still holds for $\nu > 1$, as long as $\alpha$ remains approaching zero.  Replacing the effective temperature by 
\begin{eqnarray}
&&\tilde{T}_L = \frac{T_L}{\delta} = \frac{\nu^2+3}{8\pi\nu},\nonumber\\
&&\tilde{T}_R = \frac{T_R}{\delta} = \frac{(\nu^2+3)(r_+-r_-)}{8\pi\left[\nu(r_++r_-)-\sqrt{r_+r_-(\nu^2+3)}\right]},\nonumber\\
&& {\rm with} \quad \delta \equiv \nu(r_++r_-)-\sqrt{r_+r_-(\nu^2+3)}, 
\end{eqnarray}  
we are able to obtain the same greybody factor as before.  We would like to point out that this observation is consistent with the results obtained by commutating the quasinormal modes in Ref. \cite{Chen:2009rf}\footnote{A alternative viewpoint is that $T_{L,R}$ remains intact but instead $\omega$ is rescaled to  $\tilde{\omega}=\delta \omega$  This is just a matter of convention that depends on which coordinate system is chosen.  In our convention, the effective temperature is the same as that of BTZ at $\nu=1$ limit, nevertheless the convention in \cite{Anninos:2008fx} gives rise to correct entropy with conjectured central charges.  We thank the referee for bringing up this point.}.  We remark that although we mainly focus on massless limit in computation of absorption cross section at low energy limit, the greybody factor (\ref{greybody}) can also survive for arbitrary large $m$ in the near extremal limit such that $r_+-r_- \to 0$ and $\omega \to 0$, but keeping  $\frac{\omega}{\tilde{T}_R}$ finite\footnote{This limit should be taken before $\nu \to 1$ to be taken, if desired, in order to avoid ambiguity since two limits do not commute.}.

\section{Effective CFT description}
In this section, we would like to comment on the greybody factor (\ref{greybody}) from a viewpoint of an effective CFT$_2$ on the $(\tau,u)$-plane dual to the warped AdS$_3$ space.  Having learnt from the usual AdS$_3$/CFT$_2$ dictionary, we propose a thermal correlator in the following form\footnote{We have normalized the coefficient of correlator to agree with the computation in black hole.  In generic, there could be an ambiguity for the coupling between scalar field and boundary operator, say  $\int{\phi_0{\cal O}}$.}\cite{MullerKirsten:1998mt}:
\begin{eqnarray}
&&\bigg<{\cal O}(x_+,x_-){\cal O}(0,0)\bigg>_T\sim (2h_+-1)\bigg[\frac{\pi \tilde{T}_R}{\sinh{(\pi \tilde{T}_R x_+)}}\bigg]^{2h_+}\bigg[\frac{\pi \tilde{T}_L}{\sinh{(\pi}\tilde{T}_L x_-)}\bigg]^{2h_+},\nonumber\\
&&x_\pm \equiv u \pm \tau,\nonumber\\
&&h_+ \equiv \frac{1}{2}\left[1+\sqrt{1-\frac{3(\nu^2-1)}{\nu^2}k^2+\frac{4m^2\ell^2}{\nu^2+3}}\right],
\end{eqnarray}
where $h_+$ has been derived from various aspects in the Appendix.  Now we would like to claim that a particular operator ${\cal O}_*(x_+,x_-)$ carrying the conformal dimension $h^*_+ \equiv h_+ \big|_{k=k_*}$ with\footnote{This relation is hinted by the coordinate transformation between black hole and warped AdS in the asymptotic infinity\cite{Anninos:2008fx}.  We thank the referee for pointing this out.} 
\begin{equation}\label{kw_relation}
k_* = \frac{2\nu}{\nu^2+3}\omega,
\end{equation}
might be responsible for the greybody factor of the warped AdS$_3$ black hole.  To see this, we recall that the absorption cross section is given by
\begin{eqnarray}\label{absorption_correlator}
&&\int dx_+ dx_- e^{-i\omega (x_++x_-)} \bigg<{\cal O}_*(x_+,x_-){\cal O}_*(0,0)\bigg>_T\nonumber\\
&& \sim \frac{(2h^*_+-1)\sinh\left({\frac{\omega}{2T_H}}\right)}{\omega \Gamma^2(2h^*_+)}\bigg| \Gamma\left(h^*_++i\frac{\omega}{4\pi \tilde{T}_R}\right) \Gamma\left(h^*_++i\frac{\omega}{4\pi \tilde{T}_L}\right)\bigg|^2.
\end{eqnarray}
Making use of the identity $x\Gamma(x)=\Gamma(x+1)$ and eq. (\ref{identity_1}), we are able to reproduce the result in eq. (\ref{greybody}).  To understand the relation (\ref{kw_relation}) better, we first recall the coordinates adopted in the Poincar\'{e} frame (\ref{Poincare_frame}), and then we have to introduce the energy $\omega' = 2\omega$ for $t=\tau/2$.  The relation (\ref{kw_relation}) then simply reduces to 
\begin{equation}
k=\pm \omega'
\end{equation}
for $\nu=1$.  This implies that ${\cal O}_*$ can be seen as a conventional CFT operator but in a rotating frame $(t,u)$ with speed
\begin{equation}
c \equiv \frac{\partial u}{\partial t} = \frac{\omega'}{k} = \pm \frac{\nu^2+3}{4\nu},
\end{equation} 
which reduces to unit speed for undeformed AdS$_3$.  This agrees with a known fact that for $\nu=1$ the metric (\ref{warped_AdSBH}) reduces to the BTZ metric in a rotating frame.

The key formula (\ref{absorption_correlator}) can also be understood in terms of the correlator (\ref{deformed_correlator}) obtained in the Appendix.  We may first rescale\footnote{Before the rotation, we have to rescale $u\to \frac{\nu^2+3}{2\nu}u$ in order to normalize $g_{uu}$ to unity in the Poincar\`e metric \ref{Poincare_frame}.  This is to assure that after coordinate transformation, we have $k_*(u\pm t) \to \omega u_{\pm}$.  The rescaling of $t$ is relatively not important here.} and rotate the $(t,u)$-plane into $(u_+,u_-)$-plane either clockwise (right sector) or counterclockwise (left sector):
\begin{eqnarray}
&&\bigg<  {\cal O}_*(u_+,u_-){\cal O}_*(0,0)\bigg>^L\sim \frac{e^{i\omega u_+}}{|u_-|^{2h^*_+}}, \quad \mbox{counterclockwise:} \quad(t,u)\to (u_-,u_+);\nonumber\\
&&\bigg<  {\cal O}_*(u_+,u_-){\cal O}_*(0,0)\bigg>^R\sim \frac{e^{i\omega u_-}}{|u_+|^{2h^*_+}},\quad \mbox{clockwise:} \quad(t,u)\to (u_+,u_-),\\
&& u_\pm \equiv  u \pm t/2c.\nonumber
\end{eqnarray}
Assigning temperatures $\tilde{T}_{L}$ and $\tilde{T}_{R}$ to the left and right sectors respectively, and we then define the composite thermal correlator as the product of the thermal correlators in the two sectors, i.e.
\begin{equation}
\bigg<\bigg<  {\cal O}_*(u_+,u_-){\cal O}_*(0,0) \bigg>\bigg>_T \sim (2h_+-1)e^{i\omega(u_++u_-)}\bigg[\frac{\pi \tilde{T}_R}{\sinh({\pi \tilde{T}_R u_+})}\bigg]^{2h^*_+}\bigg[\frac{\pi \tilde{T}_L}{\sinh{(\pi}\tilde{T}_L u_-)}\bigg]^{2h^*_+}.
\end{equation}
This is exactly the integrand in (\ref{absorption_correlator}), and therefore the same result is expected.  This construction can be seen as an effective CFT$_1$ coupled with a free $U(1)$ field.  This is similar to an earlier description of dimensional reduction from AdS$_3$ to AdS$_2$ for the BTZ black hole with the Chern-Simons and other higher derivative terms\cite{Sahoo:2006vz}, which has recently been extended to the warped AdS$_3$ black hole\cite{Kim:2009jm}.

\section{Superradiant Modes}
Another unusual feature for the warped AdS$_3$ black hole is the appearance of the superradiant modes when 
\begin{equation}\label{omega_bound}
\omega^2 > \frac{(\nu^2+3)^2}{12(\nu^2-1)}+ \frac{\nu^2+3}{3(\nu^2-1)}m^2\ell^2,
\end{equation}
and $\nu \neq 1$. Note that this phenomenon is particular to the warped AdS$_3$ black hole and would not occur in the $BTZ$ case where $\nu=1$.  In eqs. (\ref{phi_infty}) and (\ref{phi_decomposition}), the exponents of $r$ for the wavefunction in spatial infinity becomes complex.  Therefore, both terms of $A_{in}$ in eq. (\ref{A_coeff}) will contribute to the absorption cross section given by eq. (\ref{absorp}).  From the CFT side, the existence of the superradiant modes indicates that conformal dimension $h_{\pm}$ become complex conjugates and the {\sl effective} mass of scalar is below the Breitenlohner-Freedman bound
\begin{equation}
m_{eff}^2 \equiv \frac{4m^2}{\nu^2+3}-\frac{12(\nu^2-1)\omega^2}{(\nu^2+3)^2\ell^2} < -\frac{1}{\ell^2}.
\end{equation}
As a result, the scattering process becomes unstable.  Because of the connection in (\ref{kw_relation}), the condition for $\omega$ can be translated into a condition on $k_*$ and condition (\ref{omega_bound}) leads to a bound on angular velocity just as in the case of Kerr black holes\footnote{We remark another interesting direction towards CFT calculation of superradiance in Kerr black holes by a recent paper \cite{Bredberg:2009pv}.  The near horizon limit of extremal Kerr black holes (NHEK) also possesses the same  $SL(2,R)\times U(1)$ symmetry as in the warped AdS black holes.}.

\acknowledgments
The authors would like to thank Pei-Ming Ho, Shunsuke Teraguchi and Feng-Li Lin for helpful discussion.  This project was partially supported by the Taiwan's National Science Council and National Center for Theoretical Sciences under Grant No. NSC95-2112-M-003-013-MY3 (HCK), NSC97-2112-M-002-015-MY3 (WYW) and NSC97-2119-M-002-001.

\appendix

\section{Warped AdS/CFT correspondence}
\subsection{Warped AdS calculation}
In this section, we demonstrate the dictionary between the spatial warped $AdS_3$ and the corresponding $CFT_2$. In particular, we would like to derive the conformal weight $h_{\pm}$ from the viewpoints of spacetime isometry and boundary conformal group as well as the two-point correlator in the Poincar\`e patch\footnote{Part of discussion in this appendix has overlap with that in \cite{Anninos:2009jt}.}.  We first consider the gravity side with the following warped geometry in the global coordinate:
\begin{equation}\label{warped_AdS}
ds^2= \frac{\ell^2}{\nu^2+3}\left[ -\cosh^2{\sigma}d\tau^2+d\sigma^2+\frac{4\nu^2}{\nu^2+3}(du+\sinh{\sigma}d\tau)^2 \right].
\end{equation}
Let us now consider the Klein-Gordon equation of a massive scalar field $\Phi(\tau,\sigma,u)$ in this background.  Assuming a stationary form $\Phi=e^{-i\omega_g \tau + i k u}\phi_g(\sigma)$, the equation becomes
\begin{equation}\label{warpAdS_global}
\frac{d^2\phi_g}{d\sigma^2}+\tanh{\sigma}\frac{d\phi_g}{d\sigma} +\left[  \omega_g^2 \sech^2{\sigma}+2\omega_g k \sech{\sigma}\tanh{\sigma}-(\frac{\nu^2+3}{4\nu^2}-\tanh^2{\sigma})k^2-\frac{m^2\ell^2}{\nu^2+3} \right]\phi_g=0
\end{equation}
after separation of variables. 
It can be brought into a standard form of the hypergeometric equation
\begin{eqnarray}
&&\hskip-1cm z_g(1-z_g)f''+\big[c_g-(1+a_g+b_g)z_g \big]f'-a_g b_g f=0
\end{eqnarray}
via the following change of variable
\begin{eqnarray}
&&\phi_g = z_g^{(\omega_g+ik)/2}(1-z_g)^{(\omega_g-ik)/2}f(z_g),\nonumber\\
&&z_g\equiv \frac{1+i\sinh{\sigma}}{2}.
\end{eqnarray}
Here, prime denotes differentiation with respect to $z_g$. $a_g, b_g$ are determined by 
\begin{eqnarray}
&&a_g b_g = \omega_g(\omega_g+1)+\frac{3k^2(\nu^2-1)}{4\nu^2}-\frac{m^2\ell^2}{\nu^2+3},\qquad a_g+b_g = 1+2\omega_g,\nonumber
\end{eqnarray}
and $c_g=1+\omega_g+ik$.

On the other hand, the asymptotic behaviour of the two independent solutions to this equation is given by 
\begin{eqnarray}\label{phig}
&&\phi_g(z_g)\to C_+ z_g^{h_+} + C_- z_g^{h_-},\\
&&{\rm with} \quad h_{\pm} = \frac{1}{2}(1\pm \Delta),\qquad \Delta \equiv \sqrt{1-\frac{3(\nu^2-1)}{\nu^2}k^2+\frac{4m^2\ell^2}{\nu^2+3}},\nonumber\\
&&C_+ =(-1)^{a_g}\frac{\Gamma(c_g)\Gamma(b_g-a_g)}{\Gamma(b_g)\Gamma(c_g-a_g)},\qquad C_- =(-1)^{b_g}\frac{\Gamma(a_g-b_g)}{\Gamma(a_g)\Gamma(c_g-b_g)}. \nonumber
\end{eqnarray}
For non-integer $\Delta<1$, both solutions are renormalizable and either term in eq. (\ref{phig}) can act as a source or fluctuation.  For $\Delta>1$, only the first term is renormalizable. By setting $C_-=0$, we obtained the following quantization condition:
\begin{equation}
\omega = -(n+h_+), \qquad n = 0,1,2,\cdots.
\end{equation}
In the case of $\Delta<1$, a similar condition may be achieved
\begin{equation}
\omega = -(n+h_-), \qquad n=0,1,2,\cdots,
\end{equation}
if we choose the second term to be the renormalizable one.

\subsection{Group theory calculation}
Now let us turn to the group theory side. Here, the boundary conformal group has been deformed into $SL(2,R)_R\times U(1)_L$.  Following the notation in \cite{Anninos:2008fx}, the subgroup $SL(2,R)_R$ can be realized by
\begin{eqnarray}
&&\tilde{J}_0 = 2\partial_\tau,\nonumber\\
&&\tilde{J}_1 = 2\sin{\tau}\tanh{\sigma}\partial_\tau - 2\cos{\tau}\partial_\sigma + \frac{2\sin{\tau}}{\cosh{\sigma}}\partial_u,\nonumber\\
&&\tilde{J}_2 = -2\cos{\tau}\tanh{\sigma}\partial_\tau - 2\sin{\tau}\partial_\sigma - \frac{2\cos{\tau}}{\cosh{\sigma}}\partial_u,
\end{eqnarray}
and the additional $U(1)_L$ by
\begin{equation}
J_2=2\partial_u.
\end{equation}
It is convenient to use the following combination:
\begin{eqnarray}
&&\tilde{L}_0 = i\tilde{J}_0,\qquad \tilde{L}_+ = -(\tilde{J}_1-i\tilde{J}_2),\qquad \tilde{L}_- = \tilde{J}_1 + i \tilde{J}_2,\nonumber\\
&&L_0 = i J_2, \qquad L_+ = L_- =0.
\end{eqnarray}
The Casimirs operators are given by
\begin{equation}
L^2=\frac{1}{2}(L_+L_-+ L_-L_+)-L_0^2,\qquad \tilde{L}^2 = \frac{1}{2}(\tilde{L}_+\tilde{L}_-+ \tilde{L}_-\tilde{L}_+)-\tilde{L}_0^2.
\end{equation}
We may rewrite eq. (\ref{phig}) in terms of the d$^,$Lambertian, which can in turn be expressed in terms of the left and right Casimirs
\begin{equation}
-\frac{1}{4}(L^2+\tilde{L}^2)\Phi = \frac{m^2\ell^2}{\nu^2+3}\Phi.
\end{equation}
Solutions satisfying $L_+\Phi_+ =0$ and $\tilde{L}_-\Phi_-=0$ can be found explicitly and
\begin{equation}
\Phi_{\pm} = e^{-ih^0_{\pm}\tau} (\sech{\sigma})^{h^0_{\pm}}, \qquad h^0_{\pm}=\frac{1}{2}\left[1\pm\sqrt{1+\frac{4m^2\ell^2}{\nu^2+3}}\right].
\end{equation}
This agrees with the earlier result of $h_\pm$ for $k=0$.  To render the desired $k$-dependence, we may introduce the {\sl un}twist generators for the right-moving mode in which the $\partial_u$ term is removed:
\begin{equation}
\tilde{J}_1 \to \tilde{J}_1 - \frac{\sin{\tau}}{\cosh{\sigma}}J_2,\qquad \tilde{J}_2 \to \tilde{J}_2 + \frac{\cos{\tau}}{\cosh{\sigma}}J_2.
\end{equation}
This will lead to a natural separation between generators in $SL(2,R)$ and $U(1)_L$, which can be seen later in the computation of the correlator. 

Together with a new solution
\begin{equation}
\Phi_{\pm} = e^{-ih_{\pm}\tau + i\epsilon k u}(\sech\sigma)^{h_{\pm}},
\end{equation}
it can be shown to reproduce $h_{\pm}$ for
\begin{equation}
\epsilon^2 = \frac{3(\nu^2-1)}{4\nu^2}.
\end{equation}

\subsection{Correlator in Poincar\`e coordinate}
To obtain the correlator of the corresponding operators, it is also worthwhile to work out the same wavefunction in the Poincar\`e coordinate.  Although a general Poincar\`e coordinate valid for all range of $\sigma$ is not available as far as we know, it is sufficient to know its behaviour in the limit $\sigma\to \infty$ where a boundary CFT is conjectured to exist. The local patch can be achieved by the coordinate transformation $x=e^{-\sigma}$ and $t=\tau/2$:
\begin{equation}\label{Poincare_frame}
ds^2 \to \frac{\ell^2}{(\nu^2+3)x^2}\bigg[ \frac{3(\nu^2-1)}{\nu^2+3}dt^2 + dx^2 + \frac{4\nu^2}{\nu^2+3}x^2du^2 + \frac{8\nu^2}{\nu^2+3}xdtdu \bigg].
\end{equation}
As in Ref. \cite{Witten:1998qj}, we will solve the following equation:
\begin{equation}
\bigg( \frac{1}{\sqrt{-g}}\partial_x\sqrt{-g}g^{xx}\partial_x+\partial_ug^{uu}\partial_u-m^2 \bigg)K(x,u)=0.
\end{equation}
Using the ansatz $K(x,u) = e^{iku}\tilde{K}(x)$, we obtain
\begin{equation}
\tilde{K}(x) \sim x^{h_+},
\end{equation}
where we have chosen the operator with conformal weight $h_+$ as the renormalizable mode.  The group transformation of $SL(2,R)$ acting on the $(t,x)$-plane will map
\begin{equation}
K(x,u) \to K(x,u,t)=e^{iku}\left(\frac{x}{|x^2-t^2|}\right)^{h_+}.
\end{equation}
After performing a translation on $u\in U(1)$ and $t$, we define the bulk-to-boundary Green function as
\begin{equation}
K_b(x,u,t,u',t') \equiv e^{ik(u-u')}\left[\frac{x}{|x^2-(t-t')^2|}\right]^{h_+}.
\end{equation}
Therefore we obtain a bulk field which is determined in the following way
\begin{equation}
\phi(x,t,u) \sim \int{dt' du'} K_{b}(x,t,u,t',u') \phi_0(t',u').
\end{equation}
Its derivative in the limit $x\to 0$ is given by
\begin{equation}
\frac{\partial}{\partial x}\phi(x,t,u) \sim x^{h_+-1}\int{dt'\, du'}\frac{e^{ik(u-u')}\phi_0(t',u')}{|t-t'|^{2h_+}}.
\end{equation}
The on-shell action only gets contribution from the boundary term,
\begin{equation}
S_{eff} = \lim_{x\to 0}-\frac{1}{2}\int{dt\, du}\sqrt{-g}g^{xx}\phi \partial_x \phi \sim \frac{1}{2}\int{dt\,du\,dt'\,du'} \frac{e^{ik(u-u')}}{|t-t'|^{2h_+}}\phi_0(t,u)\phi_0(t',u').
\end{equation}
From the AdS/CFT dictionary between the effective action in the bulk and the generating function on the boundary, we have
\begin{equation}
e^{-S_{eff}(\phi)}=\bigg< e^{\int{\phi_0{\cal O}}} \bigg>.
\end{equation}
As a result, the correlator of the operators is given by
\begin{equation}\label{deformed_correlator}
\bigg< {\cal O}(t,u){\cal O}(t',u') \bigg> \sim \frac{e^{ik(u-u')}}{|t-t'|^{2h_+}}.
\end{equation}
We remark that this correlator does not take the usual form expected from the CFT$_2$.  Instead, it is a product of the correlator in CFT$_1$ and the free propagator along the $u$-direction, reflecting the breaking of symmetry from $SL(2,R)\times SL(2,R)$ to $SL(2,R)\times U(1)$ due to the deformation.

In summary, we have seen that in addition to mass $m$, spin $k$ is also involved in the definition of conformal weight $h_{\pm}$, thanks to the deformation $\nu \neq 1$.  This may be traced back to the nontrivial fibrated coordinate adopted in (\ref{warped_AdS}).

\end{document}